\documentclass{article}

\usepackage{graphicx}

\usepackage{algorithm}
\usepackage{algorithmic}

\newtheorem{theorem}{Theorem}

\title{Euclidean $k$-center Fair Clusterings}

\author{%
 {Ayano Moritaka} \hskip 10mm
  {Shin-ichi Nakano},\\
  {Kento Tanaka} \hskip 10mm 
  {Noriaki Yoshida}\\
  \\
 Gunma University, Maebashi-Shi 371-8510, Japan 
}

\date{\today}

\begin{document}
\maketitle

\begin{abstract}
Many 
approximation algorithms and heuristic algorithms
to find a fair clustering
have emerged.
In this paper
we define a new and natural variant of fair clustering problem
and design a polynomial time algorithm 
to compute an optimal fair clustering.
Let $P$ be a set of $n$ points on a plane,
and each point has a color in $C$, 
corresponding to a group.
For each color $q\in C$,
a lower bound $\ell(q)$ and 
an upper bound $u(q)$ are given.
Then we define the fair clustering problem as follows.
{\it The fair $k$-clustering problem}
is to
find
a partition of $P$
into a set of $k$ clusters
with a minimum cost
such that
each cluster
contains at least $\ell(q)$ and at most $u(q)$
points in $P$ with color $q$.
By $\ell(q)$ and $u(q)$
each cluster cannot contain too few or too many
points with a specific color.
If we regard a color to a gender 
or a minority ethnic group,
the clustering corresponds to a fair clustering.
\end{abstract}



\section{Introduction}
\label{sec:intro}

\begin{figure}[ht]
  \begin{center}
     \includegraphics[width=7cm,pagebox=cropbox,clip]{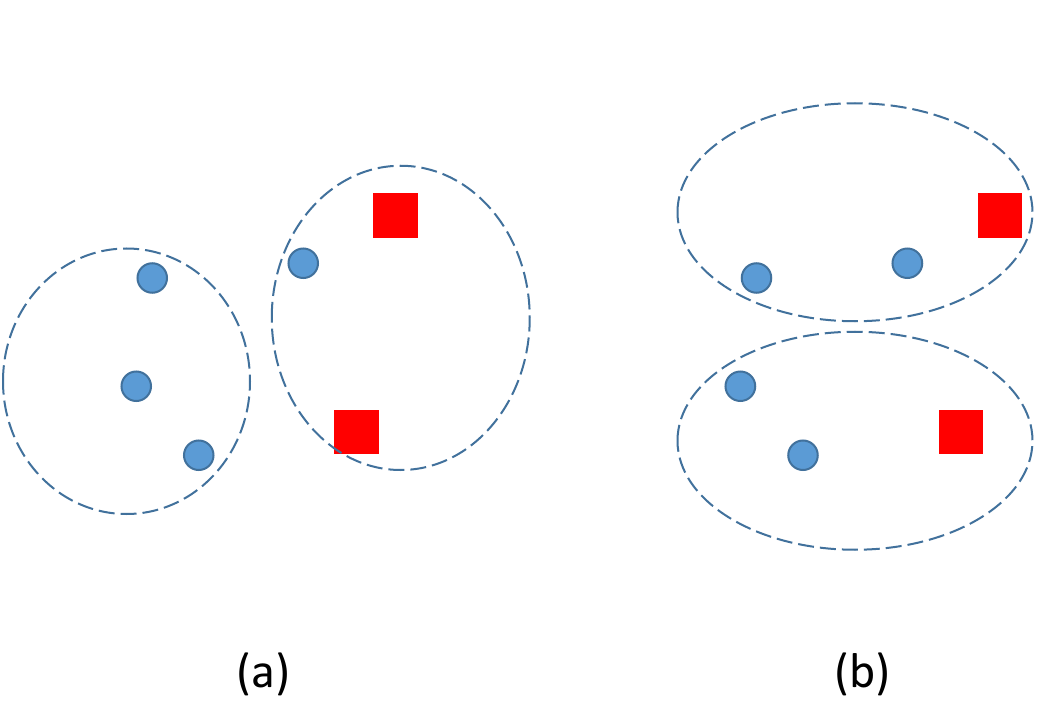}
  \end{center}
  \caption{
   (a) A non-fair clustering and
   (b) a fair clustering.
  }
  \label{fig:fig1}
\end{figure}

A clustering of a set $P$ of data is a partition of $P$
into a set of subsets of $P$, called clusters,
such that the maximum radius of the clusters is minimized.
For instance 
we may compute a clustering of home loan applicants
into several preferential interest rate clusters.
If a cluster contains too many or too few
applicants from a gender or an ethnic group,
then the clustering may not be fair, and not be desirable.
See an example in 
Fig. \ref{fig:fig1}.
Thus many research for fair clustering have emerged.
See a survey in 
\cite{C21}.

Fairness based on balance of each group
in each cluster
was first proposed 
in \cite{C17}, and
for the basic case 
with $2$ groups
and
the balance $1:t$
for some integer $t$,
a $4$-approximation algorithm 
to find a fair clustering
minimizing
the maximum radius of the clusters
is designed \cite[Theorem 13]{C17}.
It was later generalized to the 
multiple groups with more general setting
(a point can be belong several groups).
Given two parameters $\alpha$ and $\beta$, 
an approximation algorithm
to find
a fair (each cluster contains
at least $\alpha$ fraction 
and
at most $\beta$ fraction of points with each color)
clustering,
allowing a small additive violation $4\Delta +3$,
where $\Delta$ 
is the maximum number of groups a single point can be belong,
is designed
\cite{B19}.
The approximation ratio is
$(\rho + 2)$, where $\rho$ is 
the approximation ratio of
an approximation algorithm for ordinary (possibly unfair) clustering.

In this paper
we define 
a new and natural variant of the fair clustering problem
and design a polynomial time algorithm 
to compute an optimal,
not approximate,
fair clustering.
Our fairness notion is different from 
any of the twenty 
fairness notions in a survey paper\cite{C21} and a
resent paper\cite{Gu23}.

\begin{figure}[ht]
  \begin{center}
     \includegraphics[width=9cm,pagebox=cropbox,clip]{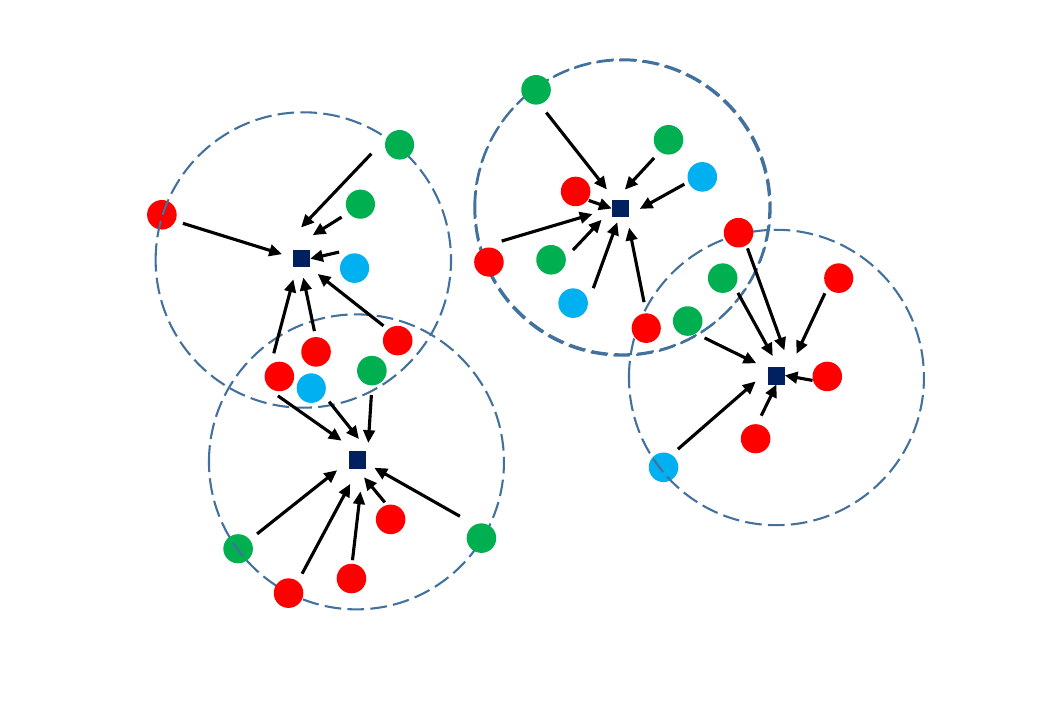}
  \end{center}
  \caption{
   A fair clustering with 
   $\ell(red)=3$, $u(red)=4$,
   $\ell(blue)=1$, $u(blue)=2$,
   $\ell(green)=2$ and $u(green)=3$.
  }
  \label{fig:fig0}
\end{figure}

Let $P$ be a set of $n$ points on a plane,
and each point has a color in $C$, 
corresponding to a group.
Also for each color $q\in C$,
a lower bound $\ell(q)$ and 
an upper bound $u(q)$ are given.
Assume $k$ is a constant (and not a part of input)
and $k$ is much smaller than $n$, say $k\le 0.01 n$.
Then {\it the fair $k$-clustering problem}
finds
a partition of $P$
into a set of $k$-clusters
such that
each cluster
contains at least $\ell(q)$ and at most $u(q)$
points in $P$ with color $q$.
See an  example in Fig. \ref{fig:fig0}.

By $\ell(q)$ and $u(q)$
a cluster cannot contain too few or too many
points with a specific color.
If we regard a color to a gender 
or a minority ethnic group,
the clustering corresponds to a fair clustering.

The rest of the paper is organized as follows.
In Section \ref{sec:def} we give some definitions
and explain an algorithm\cite{N23}
to solve the Euclidean $k$-center $(\ell,u)$-clustering problem,
which is a subroutine
in the algorithm for the fair $k$-clustering
in the next section.
%
In Section \ref{sec:algo} we give an
algorithms to compute an optimal fair $k$-clustering. 
Finally Section \ref{sec:conc} is a conclusion.

\section{Preliminaries}
\label{sec:def}

In this section we explain the algorithm to solve the $k$-center $(\ell,u)$-clustering problem in \cite{N23},
which is a subroutine in the algorithm for the fair $k$-clustering in the next Section.

Let $_xC_y$ be the number of ways to choose $y$ objects from a set of $x$ distinct objects, and
$_xC_y = x! / ((x-y)! y!) \le x^y$.


\noindent
{\bf Euclidean $k$-center $(\ell,u)$-clustering problem}

Given a set $P$ of $n$ points on a plane
and three constant integers $k$, $\ell$ and $u$,
{\it the Euclidean $k$-center $(\ell,u)$-clustering problem}
is the problem
to compute a set $Cen = \{ c_1,c_2,\cdots,c_k \}$ of $k$ center points on the plane
and
a partition $P_1\cup P_2\cup \cdots\cup P_k$ of $P$
such that

\noindent
(1) each $P_i$ contains at least $\ell$ and at most $u$
points in $P$, and

\noindent
(2) the maximum radius of $P_1, P_2, \cdots, P_k$ is minimized,
where the radius of $P_i$ is the maximum distance from $c_i$ to a point in $P_i$.

\noindent
We assume $k$ is much smaller than $n$, say $k\le 0.01 n$.

If $\ell=0$ and $u=n$,
then the problem is the ordinary
Euclidean $k$-center clustering problem,
which is intuitively to locate a set of $k$ identical disks
covering $P$ with the minimum radius.
The Euclidean $k$-center clustering problem is NP-complete
if $k$ is a part of the input \cite{M84}, but
%
if $k$, $\ell$ and $u$ are fixed integers,
a polynomial-time ($O(n^{2k+4})$ time) algorithm to solve
the Euclidean $k$-center $(\ell,u)$-clustering problem
is known \cite{N23}.
Intuitively the solution is to locate a set of $k$ identical disks
covering $P$ with the minimum radius,
plus an assignment of each point in $P$ to a center in $Cen$.

\begin{figure}[ht]
  \begin{center}
     \includegraphics[width=8cm,pagebox=cropbox,clip]{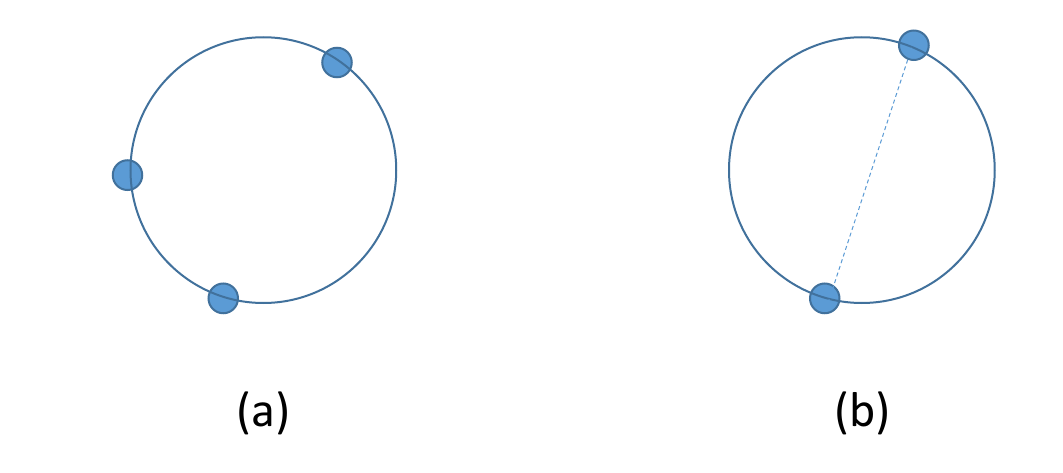}
  \end{center}
  \caption{
    (a) A disk $D$ having
three points in $P$ on the boundary, and
    (b) a disk having
two points in $P$ on the boundary
with the distance equal to the diameter.
  }
  \label{fig:fig02}
\end{figure}

The algorithm is as follows \cite{N23}.
We start with the following two observations
for the $k$ disks corresponding to the solution. 

\noindent{\bf Observation 1:}
The $k$ disks contain
a disk $D$ having
either 
(1)
three or more points in $P$ on the boundary
or
(2)
two points in $P$ on the boundary
with the distance equal to the diameter.
See Fig. \ref{fig:fig02}.

Otherwise there is a solution with less radius, a contradiction.
Thus 
the number of possible radii of the solution
is at most $_n C_3 + _n C_2 \le 2 n^3$.

\noindent{\bf Observation 2:}
Each disk in the $k$ identical disks
has either
(1) at least two points in $P$ on the boundary,
or
(2) exactly one point in $P$ on the boundary
(at the highest $y$-coordinate).


Otherwise we can move each disk so that
either (1) or (2) has satisfied
and
the set of points in $P$
covered by the disk remains the same.

\begin{figure}[ht]
  \begin{center}
   \includegraphics[width=8cm,pagebox=cropbox,clip]{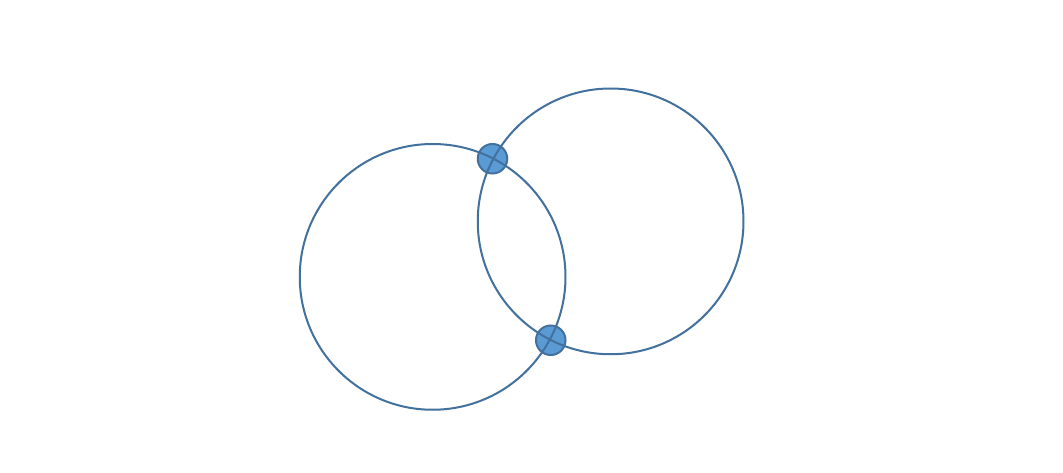}
  \end{center}
  \caption{
    Each two points in $P$
possibly define the two disks having the two points on the boundary
and with the radius same to $D$.
  }
  \label{fig:fig3}
\end{figure}

Now we explain a polynomial-time algorithms to solve
the Euclidean $k$-center $(\ell,u)$-clustering problem.
The algorithm solves
a set of the maximum flow problems with lower bounds, defined below.

Given a network $N=(V,A)$ with two vertices $s, t\in V$,
and a capacity $u(i,j)\ge 0$ and a lower bound $\ell(i,j)\ge 0$
associated with each arc $(i,j)\in A$,
a flow consisting of $f(i,j)$ for each arc $(i,j)\in A$
is {\it feasible} if
$\sum_{(i,x)\in A, x\in V} f(i,x)= \sum_{(x,i)\in A, x\in V} f(x,i)$ for each $i\in V/\{s,t\}$
and
$\ell(i,j) \le f(i,j)\le u(i,j)$ for each $(i,j)\in A$.
{\it The maximum flow problem with lower bounds}
is the problem to compute
a feasible flow with the maximum $\sum_{(s,x)\in A, x\in V} f(s,x)$.
One can solve the problem in polynomial time,
by a reduction to the ordinary maximum flow problem \cite{A93},
where the ordinary maximum flow problem is
the maximum flow problem 
with $\ell(i,j) =0$ for each $(i,j)\in A$.
The reduction needs $O(|V|+|A|)$ time.
So one can solve the maximum flow problem with lower bounds in polynomial time
using an algorithm to solve the ordinary maximum flow problem,
say in $O(|V|^3)$ time (Theorem 26.30 in \cite{C09}).

Now we explain a reduction from
the Euclidean $k$-center $(\ell,u)$-clustering problems
to a set of the maximum flow problems with lower bound.

Every two or three points in $P$ 
define a possible disk, say $D$, in Observation 1.
The number of possible $D$ is at most $2 n^3$ by Observation 1.
Then repeat the following $k-1$ times.

Either
choose other two points in $P$
possibly define 
the two disks having the two points on the boundary
and with the radius same to $D$
(See Fig. \ref{fig:fig3})
or
choose other one point in $P$
which defines
the disk having the point on the boundary
at the highest $y$-coordinate
and with the radius same to $D$.
The number of possible disks is 
at most $2(n-2)^2+n \le 2n^2$

Thus 
the number of the sets of $k$ disks
possibly corresponding to a solution of
the Euclidean $k$-center $(\ell,u)$-clustering problem is
at most $2 n^3 (2n^2)^{k-1}$.

For each set of such possible $k$ disks $\{D_1,D_2,\cdots,D_k\}$
with the same radius,
let
$Cen=\{c_1,c_2,\cdots,c_k \}$ be the centers
of the $k$ disks,
then
we check if the set of $k$ disks actually correspond to a solution of
the Euclidean $k$-center $(\ell,u)$-clustering problem or not,
as follows.

\begin{figure}[ht]
  \begin{center}
     \includegraphics[width=9cm,pagebox=cropbox,clip]{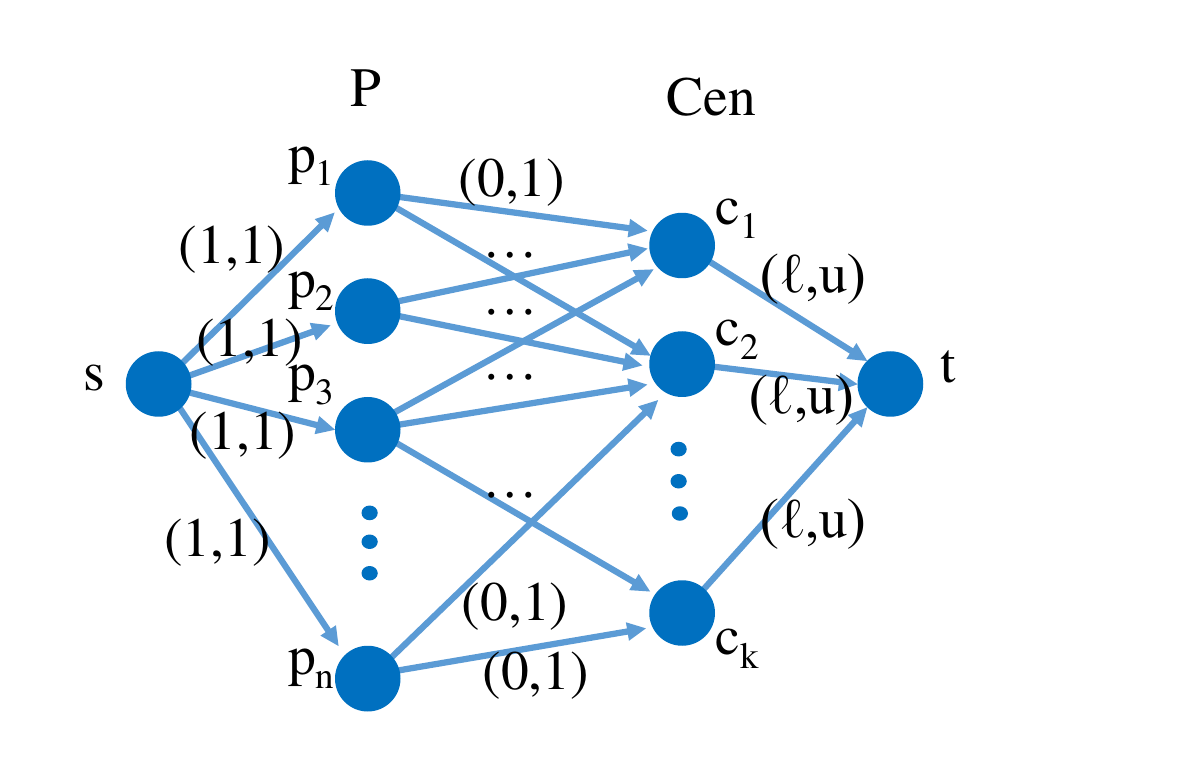}
  \end{center}
  \caption{
   A network derived from a Euclidean $k$-center 
   $(\ell,u)$-clustering problem.
  }
  \label{fig:fig03}
\end{figure}

First
we check if the $k$ disks cover $P$ or not in $O(kn)$ time.
If the $k$ disks cover $P$,
then
we construct the following instance of
the maximum flow problem with lower bounds
(See Fig. \ref{fig:fig03})
in $O(kn)$ time, and solve it in $O(n^3)$ time.

\noindent
Let $N=(V,A)$, where

\noindent
$V = \{s\} \cup P \cup Cen\cup \{t\}$, and

\noindent
$A = \{(s,p)| p\in P \} \cup $
$\{(p,c_1)| D_1$ contains
$p\in P  \} \cup $\\
$\{(p,c_2)| D_2$ contains
$p\in P  \} \cup \cdots \cup \{(p,c_k)| D_k$
contains
$p\in P  \} \cup $
$\{ (c,t)| c\in Cen \}$.


\noindent
Set
$\ell(s,p) = 1$ and $u(s,p) = 1$ for each $(s,p)\in A$ with $p\in P$,

\noindent
$\ell(p,c) = 0$ and $u(p,c) = 1$
for each $(p,c)\in A$ with $p\in P$ and $c\in Cen$,

\noindent
$\ell(c,t) = \ell$ and $u(c,t) = u$
for each $(c,t)\in A$ with $c\in Cen$.

\vskip 2mm

If the maximum flow problem with lower bounds has a solution $f$,
then
it generates a partition of $P=P_1\cup P_2\cup \cdots \cup P_k$
where $P_i = \{p | f(p,c_i) =1 \}$,
and this partition 
and the set $\{c_1,c_2,\cdots,c_k\}$
of centers
corresponds to a possible solution of
the $k$-center $(\ell,u)$-clustering problem.
Note that each $P_i$ has at least $\ell$ and at most $u$ points in $P$
since $\ell(c_i,t)=\ell$ and $u(c_i,t)=u$ for each $i$.

One can solve each maximum flow problem with lower bounds above
in $O(n^3)$ time (Theorem 26.30 in \cite{C09}),
since $|V|=1+n+k+1$ and $k$ is a constant.

The number of the maximum flow problems with lower bounds
is at most $2 n^3\cdot (2n^2)^{k-1}$, 
so
the total time to solve the problems is
$O(2 n^3\cdot (2n^2)^{k-1}\cdot n^3)$
$= O(n^{3+2(k-1)+3})$
$= O(n^{2k+4})$.

Thus we have the following theorem\cite{N23}.

\begin{theorem}\label{th:main2} 
One can solve the $k$-center $(\ell,u)$-clustering problem on a plane
in $O(n^{2k+4})$ time.
\end{theorem}

\section{Algorithm}
\label{sec:algo}

In this Section
we design an algorithm
to compute an optimal solution of the fair $k$-clustering
problem.

For each set $S$ consisting of possible $k$ disks
with the same radius
(so we regard that
$S$ consists of $k$ of center points and the common radius),
we check the following.

If $S$ covers $P$ then we proceed.
For each color $q\in C$
let $P_q$ be the set of points with color $q$.
For every $q$,
if $P_q$ has a solution of
the Euclidean $k$-center 
$(\ell(q),u(q))$-clustering
corresponding to $S$,
then
$S$ is a candidate of a solution
of the fair clustering.
Otherwise,
for some $q$,
$P_q$ has no solution of
the Euclidean $k$-center 
$(\ell(q),u(q))$-clustering
corresponding to $S$,
then
$S$ is not a solution
of the fair clustering.

Finally
choose the $S$ with the minimum radius.

The algorithm is as follows.

\begin{algorithm}[H]
	\caption{Fair Clustering$(P, C, k)$}
	\label{alg1}
\begin{algorithmic}[1]
\STATE ANS=NIL
\FOR{each set $S$ of possible $k$ disks}
   \IF{$S$ covers $P$} 
        \STATE FLAG = YES
        \FOR{each color $q\in C$}
           \STATE /* Let $P_q$ be the set of points with color $q$ */     
            \IF{$P_q$ has no solution of the Euclidean $k$-center $(\ell(q),u(q))$-clustering corresponding to $S$}
            \STATE FLAG = NO
            \ENDIF
        \ENDFOR
        \IF{FLAG = YES and $S$ has a smaller common radii than ANS}
            \STATE /* update the best ANS */ 
            \STATE ANS = S
        \ENDIF
    \ENDIF
\ENDFOR
\STATE return ANS

\end{algorithmic}
\end{algorithm}

The number of possible $k$ disks is
at most 
$O(2 n^3\cdot (2n^2)^{k-1})=$
$O(n^{2k+1})$,
and we need $O(kn)$ time to check if $S$ cover $P$,
and 
$O(n^3)$ time to solve  
a $(\ell(q),u(q))$-clustering problem
for each $q\in C$,
so
we need
$O(|C|n^{2k+4})$ time in total.

Thus we have the following theorem.

\begin{theorem}\label{th:main}
One can compute an optimal solution of a fair clustering problem on a plane
in $O(|C|n^{2k+4})$ time.
\end{theorem}

\section{Conclusion}
\label{sec:conc}

In this paper
we have defined a new and natural variant of
fair clustering problem and
designed a polynomial-time algorithm
to compute an optimal solution 
of a fair clustering problem on a plane.
The running time is
$O(|C|n^{2k+4})$.
Our algorithm works
even if a point has multiple colors.

Can we design a faster algorithm?
Our algorithm shows a theoretical complexity
(polynomial time)
of the problem, but
it is too slow for practical use.
Can we modify the algorithm with heuristics?
For instance,
one can examine a large number  (but not all ) of
possible $k$ disks chosen at random,
or
one can examine
with a descendant subsequence of possible radii.

Can we generalize the algorithm to a higher dimension?
In practical applications, this generalization is desirable.

Can we formalize the fairness of the clustering
in other way?
For instance,
let $P$ be a set of $n$ points on a plane and 
each point has a color in $C$,
then 
can we
find
a set $Cen = \{ c_1,c_2,\cdots,c_k \}$ of $k$ center points on the plane
and
an assignment of each point in $P$ to a center point  in $Cen$ 
such that
for each color $q\in C$
the maximum length from a point with color $q$
to the assigned center point
is fair (almost the same).
This means that for each color (ethnic group) $q$
the maximum distance from a point with color $q$
to the assigned facility (evacuation shelter)
is the same, so it is fair.



\small


\bibliography{ref}
\bibliographystyle{plainurl}


\vskip 2mm
Ver.
\today\quad

\end{document}